\documentclass[12pt]{iopart}

\usepackage{graphicx}

\newcommand{\beq}{\begin{equation}}
\newcommand{\eeq}{\end{equation}}
\def\Lsim{\;\raisebox{-.6ex}{$\stackrel{<}{\sim}$}\;}
\def\Gsim{\;\raisebox{-.6ex}{$\stackrel{>}{\sim}$}\;}
\newcommand{\optbar}[1]{\shortstack{{\tiny (\rule[.4ex]{1em}{.1mm})} 
  \\ [-.7ex] $#1$}}
\newcommand{\loptbar}[1]{\shortstack{{\tiny (\rule[.4ex]{1.8em}{.1mm})} 
  \\ [-.7ex] $#1$}}
\newcommand{\Loptbar}[1]{\shortstack{{\tiny (\rule[.4ex]{3em}{.1mm})} 
  \\ [-.7ex] $#1$}}
\newcommand{\Dma}{\Delta m^2_{\mathrm{atm}}}
\newcommand{\dcp}[1]{\Delta_{CP}(#1)}

\begin{document}

\title[Nu Superbeams and Factory 2002]{Neutrino Physics, Superbeams, and the Neutrino Factory\footnote{To appear in the Proceedings of the 4th International Workshop on Neutrino Factories (Nu Fact 02)}}

\author{Boris Kayser}

\address{Fermilab, MS 106, P.O. Box 500, Batavia IL 60510  USA}

\begin{abstract}
We summarize what has been learned about the neutrino mass spectrum and neutrino mixing, identify interesting open questions that can be answered by accelerator neutrino facilities of the future, and discuss the importance and physics of answering them.
\end{abstract}



\maketitle

\section{Introduction}

The impressive strength of the evidence that neutrinos can change from one flavor to another is summarized in \Tref{t1}.
\begin{table}[hbt]
\begin{center}
\caption{The strength of the evidence for neutrino flavor change. The symbol $L$ denotes the distance travelled by the neutrinos.}
\label{t1}
\begin{tabular}{@{}ll}
\br
Neutrinos & Evidence for Flavor Change\\
\mr
Atmospheric & Compelling\\
Accelerator ($L=250\,$km) & Interesting\\
Solar & Compelling\\
Reactor ($L\sim 180\,$km) & Very Strong\\
From Stopped $\mu^+$ Decay (LSND) & Unconfirmed\\
\br
\end{tabular}
\end{center}
\end{table}
Barring exotic possibilities, neutrino flavor change implies neutrino mass and mixing. Thus, neutrinos almost certainly have nonzero masses and mix. We would like to determine and then understand the spectrum of neutrino mass eigenstates $\nu_i$, each with its mass $m_i$. Why are the neutrino masses so small, and why does the neutrino mass spectrum have whatever character it turns out to have? 
Are Majorana masses involved, so that each mass eigenstate is identical to its antiparticle? We would also like to determine and then understand the character of the leptonic mixing matrix $U$, whose element $U_{\alpha i}$ describes the coupling of the charged lepton of flavor $\alpha,\; \ell_\alpha$, and the neutrino mass eigenstate $\nu_i$, to the $W$ boson. (Here, $\ell_e$ is the electron, $\ell_\mu$ the muon, and $\ell_\tau$ the $\tau$.) 
We already know that $U$, unlike its quark counterpart, contains large mixing angles. Are underlying symmetries responsible for these large mixings? And why is the leptonic mixing matrix so different from its quark counterpart? Does the leptonic mixing matrix contain CP-violating phases, leading to CP violation in the interactions of leptons? If this present-day leptonic CP violation exists, was there related leptonic CP violation in the early universe? Did the latter CP violation lead to leptogenesis---the production of unequal numbers of leptons and their antiparticles---and through this leptogenesis to the present baryon-antibaryon asymmetry in the universe?

If the neutrino flavor change reported by the LSND experiment in Los Alamos is confirmed by the MiniBooNE experiment now running at Fermilab, then either nature contains at least four mass eigenstates $\nu_i$, one linear combination of which is a sterile neutrino, or else neutrino masses violate CPT. Assuming CPT invariance, it is interesting to compare the candidate four-neutrino spectra that can accommodate all of the existing neutrino flavor change data to the new cosmological upper bound on neutrino mass. 
Combining the just-announced cosmic microwave background results from the Wilkinson Microwave Anisotropy Probe (WMAP) with earlier astrophysical results, it is found that at 95\% CL\cite{ref1},
\beq
\sum_i m_i < 0.71 \mathrm{ eV}  .
\label{eq1}
\eeq
Here, the sum runs over the masses of all the neutrino mass eigenstates. Every four-neutrino spectrum that is not strongly excluded by the flavor change data is at least somewhat challenged by the bound of \Eref{eq1} \cite{ref2}. However, this bound depends on cosmological assumptions \cite{ref3}. In addition, if CPT is broken \cite{ref4}, all data, including the new bound of \Eref{eq1}, can be accommodated \cite{ref5}. Thus, the question of whether the LSND flavor change, with its far-reaching implications, is genuine remains an issue that must be settled experimentally. To accomplish this goal is the aim of MiniBooNE.

\begin{figure}[htb]
\begin{center}
\includegraphics[width=14cm]{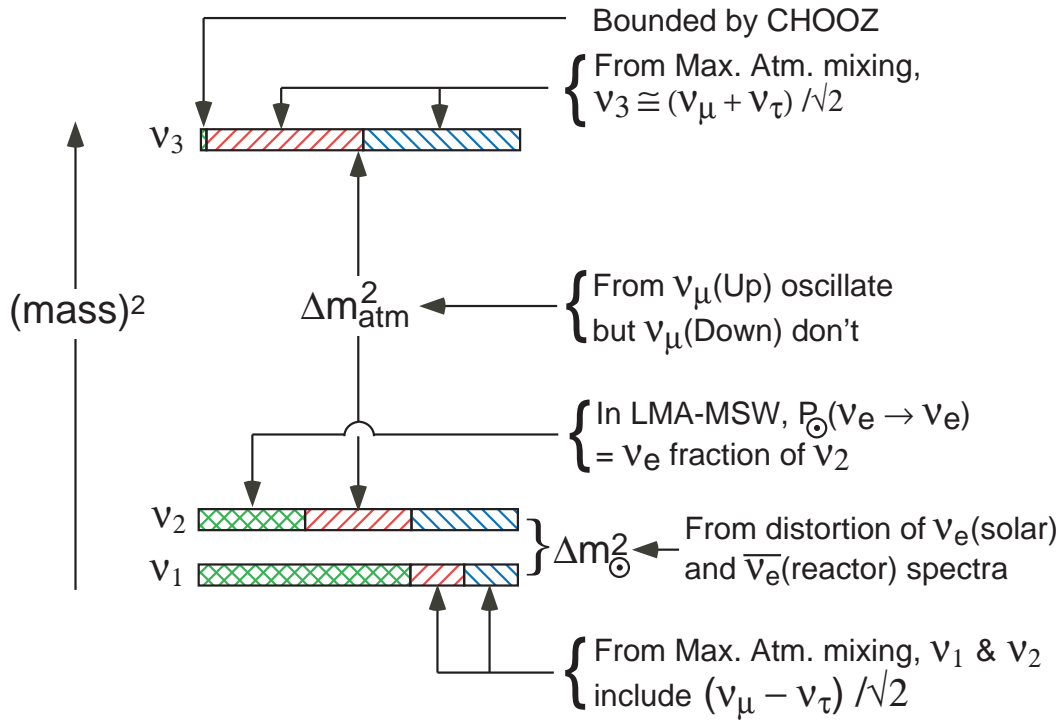}
\end{center}
\caption{A three-neutrino (mass)$^2$ spectrum that accounts for the flavor changes of the solar and atmospheric neutrinos. The $\nu_e$ fraction of each mass eigenstate is crosshatched, the $\nu_\mu$ fraction is indicated by right-leaning hatching, and the $\nu_\tau$ fraction by left-leaning hatching.}
\label{fig1}
\end{figure}

If the LSND flavor change turns out not to be genuine, then perhaps nature contains only three neutrino mass eigenstates. In that case, given all that we have learned from neutrino flavor change---that is, neutrino ``oscillation''---experiments, the neutrino spectrum is as shown in \Fref{fig1}. 
It contains a pair of mass eigenstates, $\nu_1$ and $\nu_2$, separated in (mass)$^2$ by the splitting $\Delta m^2_\odot \equiv m^2_2 - m^2_1 \sim 7 \times 10^{-5}\,$eV$^2$ that drives solar neutrino flavor change. (The symbol $\odot$ is the astronomer's symbol for the sun.) In addition, the spectrum contains a third mass eigenstate, $\nu_3$, separated from the ``solar pair'' by the much larger splitting $\Delta m^2_{\mathrm{atm}} \equiv m^2_3 - m^2_2 \sim 2.5 \times 10^{-3}\,$eV$^2$ that drives atmospheric neutrino flavor change. The isolated neutrino $\nu_3$ might be above the solar pair, as shown in \Fref{fig1}, or below it.

The neutrino state of flavor $\alpha$, which is the state $|\nu_\alpha \rangle$ produced in association with the charged lepton $\ell_\alpha$ of flavor $\alpha$ in the leptonic decay $W^+ \to \ell_\alpha^+ + \nu_\alpha$, is a superposition of the mass eigenstates $\nu_i$. This superposition is given by 
\beq
|\nu_\alpha \rangle = \sum_i U^*_{\alpha i} |\nu_i \rangle  .
\label{eq2}
\eeq
Correspondingly, each mass eigenstate $\nu_i$ is a superposition of the neutrinos $\nu_\alpha$ of definite flavor. Using the unitarity of the mixing matrix $U$, it follows from \Eref{eq2} that this superposition is given by 
\beq
|\nu_i \rangle = \sum_\alpha U_{\alpha i} |\nu_\alpha \rangle  .
\label{eq3}
\eeq
Thus, the $\nu_\alpha$ fraction of mass eigenstate $\nu_i$ is $|U_{\alpha i}|^2$. The flavor fractions of each mass eigenstate are indicated in \Fref{fig1} by the hatching code explained in the caption. The significance of these fractions is the following: If, for example, $\nu_2$ interacts in some target and produces a charged lepton, the probability that this charged lepton is an e is the $\nu_e$ fraction of $\nu_2$, the probability that it is a $\mu$ is the $\nu_\mu$ fraction, and the probability that it is a $\tau$ is the $\nu_\tau$ fraction.

\Fref{fig1} summarizes how we have learned the flavor contents of the various mass eigenstates, and the (mass)$^2$ splittings between them. Let us explain how these features of the neutrino spectrum were found, starting at the top.

The $\nu_e$ fraction of $\nu_3,\;|U_{e3}|^2$, is known to be $\Lsim$ 0.03 at 90\% CL from the bound on Short Base Line reactor $\overline{\nu_e}$ oscillation \cite{ref6}. The reactor experiments that produced this bound would have been sensitive to $\overline{\nu_e}$ disappearance through oscillation into other flavors with an oscillation frequency of $\Delta m^2_{\mathrm{atm}}$. 
The fact that they did not observe any such disappearance implies that $\nu_3$, at one end of the $\Delta m^2_{\mathrm{atm}}$ (mass)$^2$ gap, does not couple appreciably to an electron, or else does not couple appreciably to the other charged leptons, the muon and tau. In the latter case, $\nu_3$ would be almost entirely $\nu_e$, but this is impossible, since the members of the solar pair, $\nu_1$ and $\nu_2$, must have appreciable $\nu_e$ fractions in order to carry out their role in the evolution of solar electron neutrinos. 
By the unitarity of the $U$ matrix, the $\nu_e$ fractions of $\nu_1,\; \nu_2$ and $\nu_3$ must add up to unity: $\sum_i |U_{ei}|^2 = 1$. Thus, the $\nu_e$ fraction of $\nu_3$ cannot be close to unity, so that, given the reactor bound, it must be close to zero.

With the $\nu_e$ fraction of $\nu_3$ small, this mass eigenstate is almost entirely $\nu_\mu$ and $\nu_\tau$. Given its position at one end of the atmospheric (mass)$^2$ gap, it obviously plays a major role in the oscillation of the atmospheric neutrinos. The latter oscillation is very well described by the hypothesis that $\nu_\mu \to \nu_\tau$ with mixing that is very large and possibly maximal \cite{ref7}. Maximal mixing would mean that, apart from its small $\nu_e$ component, $\nu_3$ is a 50-50 mixture of $\nu_\mu$ and $\nu_\tau:\; \nu_3 \cong (\nu_\mu + \nu_\tau)/\sqrt{2}$.

The very rough size of $\Delta m^2_{\mathrm{atm}}$ follows from the fact that when matter effects are negligible, the probability for neutrino oscillation depends on the relevant $\Delta m^2$, on the distance $L$ that the neutrinos travel, and on their energy $E$, according to $\sin^2 [1.27 \Delta m^2 (\mathrm{eV})^2\, L(\mathrm{km}) / E(\mathrm{GeV})]$. 
The atmospheric neutrinos of energy $\sim\,$1GeV are observed to oscillate appreciably when $L$ is the diameter of the earth ($\sim 10^4\,$km), but insignificantly when $L$ is only the thickness of the lower atmosphere ($\sim 10$ km). Thus $10^{-4} \Lsim \Delta m^2_{\mathrm{atm}} \Lsim 10^{-2}$ eV$^2$. Detailed studies of atmospheric neutrino oscillation vs. $L$ and $E$ have yielded a much more precise $\Delta m^2_{\mathrm{atm}} \sim 2.5 \times 10^{-3}$ eV$^2$.

The solar neutrino data, the KamLAND Long Base Line reactor antineutrino results \cite{ref8}, and the analyses of all these data make it extremely likely that the Large Mixing Angle version of the Mikheyev-Smirnov-Wolfenstein effect (LMA-MSW) is responsible for the behavior of the solar neutrinos. For the high-energy solar neutrinos from $^8$B decay, the approximate result of LMA-MSW is that each of these neutrinos arrives at the surface of the earth as a $\nu_2$, the heavier member of the solar pair of mass eigenstates. 
The probability that a $^8$B solar neutrino, born as a $\nu_e$, will be detected as a $\nu_e$, rather than as a neutrino of another flavor, at the surface of the earth is then just the $\nu_e$ fraction of $\nu_2$. This probability is observed to be $\sim 1/3$ \cite{ref9}, so the $\nu_e$ fraction of $\nu_2,\; |U_{e2}|^2$, must be $\sim 1/3$. Since $\sum |U_{ei}|^2 = 1$, and $|U_{e3}|^2 \Lsim 0.03$, the $\nu_e$ fraction of $\nu_1,\; |U_{e1}|^2$, must be $\sim 2/3$.

Assuming maximum mixing in the atmospheric $\nu_\mu \to \nu_\tau$ oscillation, each of $\nu_2$ and $\nu_1$ contains $\nu_\mu$ and $\nu_\tau$ in equal proportion for the same reason that $\nu_3$ does. With maximum $\nu_\mu - \nu_\tau$ mixing, the $\nu_{\mu,\tau}$ portion of either $\nu_2$ or $\nu_1$ is the maximally mixed neutrino $(\nu_\mu - \nu_\tau)/\sqrt{2}$ that is orthogonal to the one [$(\nu_\mu + \nu_\tau)/\sqrt{2}$] found in $\nu_3$. 
Of course, $\sim 1/3$ of $\nu_2$ is taken up by its $\nu_e$ fraction, so $(\nu_\mu - \nu_\tau)/\sqrt{2}$ occurs in $\nu_2$ with a coefficient reducing its size to the remaining $\sim 2/3$ of this neutrino, and similarly for $\nu_1$.

The solar pair $\nu_2$ and $\nu_1$, with their large couplings to the electron, govern the evolution of both solar $\nu_e$ and reactor $\overline{\nu_e}$. Information on the splitting $\Delta m^2_\odot$ between these mass eigenstates has come from both the energy dependence of the survival probability of solar $\nu_e$ (including low-energy $\nu_e$ from the solar pp fusion chain) and the spectral and rate observations of the KamLAND reactor experiment.

The flavor fractions $|U_{\alpha i}|^2$ shown in \Fref{fig1} do not indicate the signs and possible CP-violating phases of the elements $U_{\alpha i}$. To convey this information, we display $U$. Assuming maximum mixing in the atmospheric neutrino oscillation, $U$ is given approximately by \cite{ref10}
\begin{eqnarray}  
  &  &  \hbox{\hskip1.7cm}\nu_1  \hbox{\hskip2.5cm} 
       \nu_2  \hbox{\hskip1.8cm} \nu_3     \nonumber  \\
 U  &  = &  \begin{array}{c}
 	\nu_e  \\  \nu_\mu \\ \nu_\tau 
 		\end{array}
\hspace{-0.25cm}\left[ 
\begin{array}{ccc}
	\phantom{-}c\,e^{i\alpha_1/2} 
         & \phantom{-}s\,e^{i\alpha_2/2} & s_{13}\,e^{-i\delta} \\
    -s\,e^{i\alpha_1/2}/\sqrt{2} 
         & \phantom{-}c\,e^{i\alpha_2/2}/\sqrt{2} & 1/\sqrt{2} \\
    \phantom{-}s\,e^{i\alpha_1/2}/\sqrt{2} 
         & -c\,e^{i\alpha_2/2}/\sqrt{2} & 1/\sqrt{2} \\ 
\end{array}
\right] ~~.
\label{eq4} 
\end{eqnarray}
Here, the symbols outside the matrix label its rows and columns, and $\nu_3$ is the isolated mass eigenstate, whether it is heavier or lighter than the solar pair. The symbols $c$ and $s$ stand for $\cos\theta_\odot$ and $\sin \theta_\odot$, respectively, where $\theta_\odot$ is the large solar mixing angle deduced from the LMA-MSW description of the evolution of solar neutrinos. At 90\% confidence level \cite{ref11},
\beq
0.25 \Lsim \sin^2 \theta_\odot \Lsim 0.40 ~~ .
\label{eq5}
\eeq
The symbol $s_{13}$ stands for $\sin \theta_{13}$ where $\theta_{13}$ is a small mixing angle describing the small $U_{e3}$. As already mentioned, at 90\% CL $\sin^2 \theta_{13} =  |U_{e3}|^2 \Lsim 0.03$. Finally, $\delta, \alpha_1$, and $\alpha_2$ are possible CP-violating phases, and are presently completely unknown. The phase $\delta$, if present, would in general lead to a CP-violating difference
\beq
\Delta_{CP}(\alpha\beta) \equiv P(\nu_\alpha \to \nu_\beta) - P(\overline{\nu_\alpha} \to \overline{\nu_\beta}) 
\label{eq6}
\eeq
between the probability $P(\nu_\alpha \to \nu_\beta)$ for the neutrino oscillation $\nu_\alpha \to \nu_\beta$, and the probability $ P(\overline{\nu_\alpha} \to \overline{\nu_\beta})$ for the corresponding antineutrino oscillation $\overline{\nu_\alpha} \to \overline{\nu_\beta}$. However, from \Eref{eq4} we see that all effects of $\delta$ are proportional to $\sin \theta_{13}$. 
Therefore, it is very important to show that $\theta_{13}$ is nonvanishing, so that CP violation in neutrino oscillation can occur, and to gain an approximate knowledge of the size of $\theta_{13}$, so that one will know what kinds of experimental facilities are needed to find the CP violation and then to study it. The phases $\alpha_{1,2}$ do not affect neutrino oscillation, but can influence the rate for neutrinoless double beta decay.

What we have learned about neutrinos in the last five years is impressive. But it is just the beginning of the unraveling of the nature of neutrinos and their interactions. We still do not know how many neutrino mass eigenstates there are. If there prove to be more than three, what does the spectrum look like? If there turn out to be only three, is the solar pair at the bottom of the spectrum (a ``normal'' spectrum) or at the top (an ``inverted'' spectrum)? 
How high above zero does this entire (mass)$^2$ spectrum lie? Do the mass eigenstates $\nu_1,\; \nu_2$, and $\nu_3$ contain $\nu_\mu$ and $\nu_\tau$ in exactly equal proportion, reflecting truly maximum $\nu_\mu - \nu_\tau$ mixing? Or, is there a deviation from maximality? If there is a symmetry behind the nearly maximum mixing, then the deviation from maximality would tell us the size of the symmetry breaking amplitude, relative to that of the $\nu_\mu - \nu_\tau$ mixing amplitude \cite{ref12}. Is each mass eigenstate $\nu_i$ a Majorana particle $(\overline{\nu_i} = \nu_i)$, as expected from the ``see-saw'' explanation of the lightness of neutrinos and other arguments? 
Is $\theta_{13}$ nonvanishing, as required for CP violation in neutrino oscillation? If so, how large is it? Does neutrino oscillation actually violate CP? What is the origin of neutrino flavor physics?

The potent accelerator neutrino facilities of the future---neutrino superbeams and a neutrino factory---will have a particularly important impact with respect to three of these questions:
\begin{enumerate}
\item How big, at least approximately, is $\theta_{13}$?
\item If there are only three neutrinos, is the solar pair at the bottom or the top of the spectrum?
\item Does neutrino oscillation violate CP?
\end{enumerate}
The expected reach of various facilities with respect to these questions is discussed in other papers in this volume. Here we shall focus on the importance to physics of the questions, and on the physics of answering them.

Demonstrating that $\theta_{13}$ is nonvanishing and determining its order of magnitude is the first goal, since CP violation in oscillation and our ability to determine whether the three-neutrino spectrum is normal or inverted both depend on $\theta_{13}$. How large do we expect $\theta_{13}$ to be? While there is no firm theoretical expectation, it would not be surprising if $|U_{e3}| = \sin \theta_{13}$ were not very much smaller than the other elements of $U$. In a gauge theory, $U$ is given by
\beq
U = X_\ell X_\nu ~~,
\label{eq7}
\eeq
where $X_\ell$ and $X_\nu$ are unitary matrices, $X_\ell$ being involved in the diagonalization of the charged lepton mass matrix, and $X_\nu$ in that of the neutrino mass matrix. From \Eref{eq7},
\beq
U_{\alpha i} = \sum_j (X_\ell)_{\alpha j} (X_\nu)_{ji} ~~.
\label{eq8}
\eeq
Now, as summarized by Equations~(\ref{eq4}) and (\ref{eq5}), all $U_{\alpha i}$ other than $U_{e3}$ are quite appreciable fractions of unity, their maximum size. From \Eref{eq8}, we see that this implies that, in general, quite a few of the elements of both $X_\ell$ and $X_\nu$ must be appreciable fractions of unity as well. 
But then, it would take a special cancellation among the terms in the sum $\sum_j (X_\ell)_{ej} (X_\nu)_{j3} = U_{e3}$ for $U_{e3}$ to be tiny compared to unity. To be sure, there could be a symmetry that causes just such a cancellation, but in the absence of such a special effect, $U_{e3}$ may not be terribly much smaller than the other elements of $U$.

Experimentally, the value of $\theta_{13}$ may be sought using the fact that, neglecting matter effects, 
\beq
\fl P(\optbar{\nu_e} \to \optbar{\nu_\mu}) \cong P(\optbar{\nu_\mu} \to \optbar{\nu_e}) \cong \frac{1}{2} \sin^2 2\theta_{13} \sin^2 \left[ 1.27 \Dma (\mathrm{eV}^2) \frac{L(\mathrm{km})}{E(\mathrm{GeV})} \right] ~~.
\label{eq9}
\eeq
One may study the small probabilities $P(\optbar{\nu_\mu} \to \optbar{\nu_e})$ using $\optbar{\nu_\mu}$ superbeams. If $\theta_{13}$ turns out to be too small to be uncovered this way, then one can study the probabilities$P(\optbar{\nu_e} \to \optbar{\nu_\mu})$ in the $\optbar{\nu_e}$ beams created by the $\mu^{+ \atop (-)}$ decays in a neutrino factory.

If the neutrino spectrum contains only three states, we can determine whether the spectral pattern is normal or inverted by studying the oscillations $\optbar{\nu_e} \to \optbar{\nu_\mu}$ or $\optbar{\nu_\mu} \to \optbar{\nu_e}$ with neutrino energy $E$ high enough and neutrino baseline $L$ long enough so that matter effects are significant \cite{ref13}. Suppose, for example, that we work under neutrino factory conditions with $E \Gsim 10$ GeV and $L \Gsim 2000$ km. The oscillation probabilities are given by 
\beq
\fl P(\optbar{\nu_e} \to \optbar{\nu_\mu}) = P(\optbar{\nu_\mu} \to \optbar{\nu_e}) \cong \frac{1}{2} \sin^2 2\loptbar{\theta_M} \sin^2 \left[ 1.27 \Loptbar{\Delta m^2_M} (\mathrm{eV}^2) \frac{L(\mathrm{km})}{E(\mathrm{GeV})} \right] ~~.
\label{eq10}
\eeq
These expressions, which assume the neutrinos travel through matter of constant density, are identical in form to the corresponding expressions for oscillation in vacuum, \Eref{eq9}, but the vacuum mixing angle $\theta_{13}$ and (mass)$^2$ splitting $\Dma$ are replaced by effective mixing angles $\loptbar{\theta_M}$ and (mass)$^2$ splittings $\Loptbar{\Delta m^2_M}$ for propagation of neutrinos and antineutrinos in matter. Let us define 
\beq
\loptbar{x_M} = \;\stackrel{+}{\scriptstyle{(}-\scriptstyle{)}}
 \frac{2 \sqrt{2} G_F N_e E}{\Dma} ~~.
\label{eq11}
\eeq
Here, $G_F$ is the Fermi coupling constant, $N_e$ is the number of electrons per unit volume in the earth through which the neutrinos pass, and $\Dma$ is defined as the (mass)$^2$ of the isolated neutrino $\nu_3$ minus the (mass)$^2$ of the nearly degenerate solar pair of neutrinos $\nu_{1,2}$. It is the sign of $\Dma$ that we wish to determine. In terms of $\loptbar{x_M}$, the matter mixing angles and (mass)$^2$ splittings are given by
\beq
\sin^2 2\loptbar{\theta_M} = \frac{\sin^2 2\theta_{13}}{\sin^2 2\theta_{13} + (\cos 2\theta_{13} - \loptbar{x_M})^2}
\label{eq12}
\eeq
and
\beq
\Loptbar{\Delta m^2_M} = \Dma \sqrt{\sin^2 2\theta_{13} + (\cos 2\theta_{13} - \loptbar{x_M})^2} ~~.
\label{eq13}
\eeq
From \Eref{eq12}, we see that
\beq
\frac{\sin^2 2\theta_M}{\sin^2 2\overline{\theta_M}} \quad \cases{>1; & normal spectrum \\ <1; & inverted spectrum \\ }
\label{eq14}
\eeq
Thus, a comparison between neutrino and antineutrino oscillations would tell us whether the spectrum is normal or inverted.

This approach to determining the character of the neutrino mass spectrum is logically very similar to the approach that was used to determine that $K_L$ is heavier than $K_S$. One passed $K$ mesons through matter known as a regenerator, and beat the sign to be determined, that of $m(K_L) - m(K_S)$, against a sign that was already known, that of the regeneration amplitude. 
In the neutrino case, one will pass neutrinos through matter, and again beat the sign to be determined, that of $m^2(\nu_3) - m^2(\nu_{1,2})$, against a sign that is already known, that of the extra energy acquired by a $\nu_e$ when it undergoes coherent forward scattering from electrons.

The pattern (and size!) of CP-violating effects in neutrino oscillation depends on how many neutrino species there are. If there are only three, the pattern becomes very simple. There are three CP-violating differences $\dcp{\alpha\beta}$ of the kind defined by \Eref{eq6}, and they obey
\begin{eqnarray}
\dcp{e\mu} & = & \dcp{\mu\tau} = \dcp{\tau e} \nonumber \\
	& = & 16\,J\, k_{12} k_{23} k_{31} ~~.
\label{eq15}
\end{eqnarray}
Here,
\beq
J \equiv \Im (U^*_{e1}\, U_{e3}\, U_{\mu 1}\, U^*_{\mu 3}) \cong \frac{1}{4} \sin 2\theta_\odot \sin\theta_{13} \sin\delta
\label{eq16}
\eeq
and
\beq
k_{ij} = \sin \left[ 1.27 \Delta m^2_{ij} (\mathrm{eV}^2) \frac{L(\mathrm{km})}{E(\mathrm{GeV})} \right] ~~,
\label{eq17}
\eeq
with $m^2_{ij} = m^2_i - m^2_j$. One sees from these relations that the different CP-violating differences that can {\em in principle} be measured must all be equal, and that all of them depend, as previously noted, not just on the phase $\delta$ but also on the small mixing angle $\theta_{13}$.

To illustrate the size of the differences $\dcp{\alpha\beta}$ of \Eref{eq15}, let us suppose that $\sin^2 \theta_\odot \simeq 0.3,\; \sin^2 \theta_{13} \simeq 0.01, \sin\delta \simeq 1, \Delta m^2_\odot \simeq 7 \times 10^{-5}\,$eV$^2,\; \Dma \simeq 2.5 \times 10^{-3}$eV$^2$ and $L/E \simeq 3000\,$km/6 GeV, so that we are working at the first peak of $k_{23} \cong -k_{31}$. Then the $\dcp{\alpha\beta}$ are between 1\% and 2\%.

In practice, the probabilities for neutrino and antineutrino oscillations in matter will depend simultaneously on the CP-violating phase $\delta$, on a number of CP-conserving neutrino parameters, and on matter-induced, but not genuinely CP-violating, neutrino-antineutrino asymmetries. It will be necessary to carry out a variety of complementary experiments and to analyze them jointly to disentangle the different neutrino properties from one another and from the matter effects \cite{ref14}.

A demonstration that neutrino oscillation violates CP would be particularly interesting for at least two reasons. First, it would establish that CP violation is not a peculiarity of quarks. Secondly, by showing that leptonic interactions do violate CP, it would raise the probability that the observed baryon asymmetry in the universe grew out of a lepton asymmetry. 
The latter would have arisen through CP violation in the decay of very heavy Majorana neutral leptons $N$, which in the see-saw mechanism for neutrino mass are heavy partners of the obvserved light neutrinos \cite{ref15}. Subsequently, the lepton asymmetry would have been converted into a baryon asymmetry by nonperturbative Standard Model processes.

The compelling evidence for neutrino mass and mixing opens a whole new world for us to explore. What we have learned about this world so far has led to very important open questions that need to be answered through future experiments. Mounting these experiments will entail the solving of interesting technical challenges, with a large scientific payoff. The coming years will be an exciting time in neutrino physics.

\ack
It is a pleasure to thank the Kavli Institute for Theoretical Physics in Santa Barbara for its warm hospitality during the writing of this paper. The author is grateful to Susan Kayser for her crucial role in the production of this manuscript.

\section*{References}

\def\plb#1#2#3{20#2 {\it Phys.\ Lett.} {\bf B#1} #3}
\def\plbo#1#2#3{19#2 {\it Phys.\ Lett.} {\bf B#1} #3}
\def\prd#1#2#3{20#2 {\it Phys.\ Rev.} {\bf D#1} #3}
\def\prdo#1#2#3{19#2 {\it Phys.\ Rev.} {\bf D#1} #3}
\def\prl#1#2#3{20#2 {\it Phys.\ Rev.\ Lett.} {\bf #1} #3}
\def\prlo#1#2#3{19#2 {\it Phys.\ Rev.\ Lett.} {\bf #1} #3}
\def\etc{{\it et al.}}

\end{document}